# Fermi-Level Pinning and Barrier Height Control at epitaxially grown ferromagnet/ZnO/metal Schottky Interfaces for opto-spintronics applications

Mohamed Belmoubarik [1,2]*

[1] International Iberian Nanotechnology Laboratory, INL, Av. Mestre José Veiga s/n, 4715-330, Braga, Portugal

[2] Department of Electronic Engineering, Tohoku University, Sendai 890-8579, Japan

*Corresponding author: mohamed.belmoubarik@inl.int ORCID: 0000-0003-3592-1259

# Abstract

Schottky contacts (SC) at the ferromagnet/ZnO interface are good candidates for the realization and control of several semiconductor emerging magnetic phenomena such spin injection and spin-controlled photonics. In this work, we demonstrate the epitaxial growth of single-phase and wurtzite-ZnO thin films on fcc Pt/Co$_{0.30}$Pt$_{0.70}$ (111) electrodes by MBE technique. While the magnetic properties of the Pt/Co$_{0.30}$Pt$_{0.70}$ buffer remain unchanged after the ZnO growth, the electric measurements of back-to-back Schottky diodes revealed a Schottky barrier height at the metal/ZnO interfaces in the range of 590–690 meV using Cu, Pt and Co$_{0.30}$Pt$_{0.70}$ contacts. A pinning factor *S* and a charge neutrality level (CNL) $\Phi_{CNL}$ of 0.08 and 4.94 eV, respectively, were obtained indicating a strong Fermi-level pining with a CNL level that lies 0.64 eV bellow the conductance band of ZnO semiconductor. These experimental findings indicate that Co$_{0.30}$Pt$_{0.70}$/ZnO interface follows the metal-induced gap states model and can open a pathway for the realization of opto-spintronics applications such spin-LEDs.

(157 words)

**Keywords:** Zinc oxide, molecular bream epitaxy, ferromagnet alloy, Schottky barrier, Fermi-level pining.

## 1) Introduction

Zinc oxide (ZnO) – based material is widely used semiconductor in electronic and optoelectronic devices, especially for short-wavelength lights generation and detection,[1,2] because of its wide band-gap of 3.34 eV with a Wurtzite (*wz*) crystal, high exciton binding energy (60 meV compared to 25 meV for GaN)[3] at room temperature (RT), higher quantum efficiency, and amenability to wet chemical etching.[4–7] Changing the ZnO material properties by means of metal doping induces versatile physical properties, such electric, magnetic and ferroelectric,[8,9] and making of it an attractive and tool for the fabrication of low power electronic devices.[10] For example, MgZnO is one of the promising candidates as a tunneling barrier in ZnO-based hetero-junction devices due to small lattice mismatching and large band gap energy (3.34 – 4.60 eV for *wz*-$Mg_xZn_{1-x}O$ ($x \sim 0 – 55\%$) vs. 7.8 eV for MgO)[1,11].

Schottky contacts (SCs) on ZnO alloys are important and useful in many applications, including high-electron-mobility transistors (HEMTs),[12,13] metal semiconductor field-effect transistors (MESFETs),[14,15] ultraviolet photodetectors,[16,17] gas sensors,[17,18] and piezoelectric nanogenerators.[19,20] SCs on n-type ZnO, are difficult to be prepared compared to ohmic contacts because of the high donor concentration at the n-ZnO surface due to native defects, such as oxygen vacancies and/or zinc interstitial atoms. To realize high-quality SCs, oxygen vacancy-related defects at the ZnO surface must be resolved by reducing the carrier concentration at the oxide surface and removing the interfacial states. The improvement of the Schottky contacts to n-type ZnO is realized by surface treatment stages that are often added before the metal is deposited. These processes include remote O/He plasma treatment[21], KrF excimer laser irradiation[22], and chemical/physical treatments utilizing diluted $H_2O_2$ solution[23] or sulfides[24]. For instance, the improvement of SC with sulfide-treated n-ZnO[24] was attributed to the formation of a low conductive interfacial ZnS and Zn vacancies that played the role of a self-compensation of free electrons at the surface region, leading to the reduction of the net carrier concentration. Another strategy consists of the elimination of the surface electron accumulation layer by the passivation of an oxidized noble metal, such as $IrO_x$-, $PdO_x$- and $PtO_x$/ZnO SCs[17,25]. Also, the type of ZnO deposition technique also play a determinant role in controlling the SC interfacial charges, and thus improving the quality of SCs.[26] Among contact metals only Pt have a small lattice mismatch with ZnO(0001) of less than 0.5%[27] which is favorable for the preparation of high-quality SCs based with less surface treatment.[28,29] For example, Allen *et al.*[30,31] could improve the quality of ZnO SCs achieving barrier heights of up to 1.2 eV and ideality factors near 1 by the oxygen treatment prior to the deposition of noble metal contacts.

Concerning the mechanism and prediction of Schottky barrier height (SBH) of ZnO SCs with metals, the following arguments were reported. For a large database of ZnO SBHs, the relationship between ZnO SBH and the metal WF wasn't trivial as concluded by Leonard J. Brillson (2011, 2013)[17,27]. Kozuka *et al.* (2014)[28] showed that depending on the surface treatment, only few ZnO-SCs (<5%) exhibited high interface quality and lie on the Schottky-Mott limit. M. Allen and S. Durbin (2010)[25] argued that metal/ZnO SCs doesn't follow Tung's chemical bond polarization (CBP) model and found a strong correlation between SBH and metal Miedema electronegativity than with metal work function (WF) by considering the metal-induced gap states (MIGS) model. Fortunately, N R D'Amico et al. (2015)[28] reported a good agreement between the calculated SBH from first principles and the MIGS model for the Zn-terminated ZnO–metal interfaces. Therefore, the need for an experimental investigation about the relationship between SBH and the metal properties at ZnO-SCs.

On the other hand, *wz*-(Mg,Zn)O alloy was reported to be an alternative tunnel-barrier for low power magnetic tunnel junctions,[34–36] and CoPt/ *wz*-MgZnO(0001) interface has been experimentally demonstrated as a versatile tool for spintronics devices.[37,38] Hence, the suitability of CoPt/ZnO SC for the generation and spin injection with LED and spin-photonics devices, as reported Co-Pt/p-Si and CoPt/GaAs SCs.[39,40] However, to the best of our knowledge, there is no report about ZnO-SC with a ferromagnetic (FM) metal unlike other semiconductors such Si and GaAs[41–43]. In this study, we choose Pt/Co$_{0.30}$Pt$_{0.70}$ as a bottom buffer layer for the following reasons: (1) the epitaxial compatibility of fcc-Pt(111) with single crystal c-sapphire substrate at high temperatures of growth,[44] (2) the smaller lattice-mismatch of *wz*-ZnO(0001) with fcc-Pt(111) [0.4%] and fcc-Co$_{0.30}$Pt$_{0.70}$ [4.8 %] thin films compared to sapphire substrates [17.3%] or Co$_3$Pt alloys [7.2%][39], (3) the high T$_C$ above 180 °C for 100-nm-thick fcc-Co$_{0.30}$Pt$_{0.70}$ favorable for spintronics applications,[46] (4) the moderate growth temperature of epitaxial *wz*-ZnO thin films by MBE within 400–600 °C preventing the oxidation of the bottom electrode and/or interdiffusion of Co and Pt atoms inside ZnO lattices,[45] and (5) the feasibility of both spin injection and detection in light-emitting Schottky diodes with CoPt contacts operating at temperatures close to RT and a narrow effective field range to ±100 mT[39,47] and (6) the possible current-induced self-switching of perpendicular magnetization in CoPt single layer.[48]

Consequently, we hereby describe the epitaxial growth of high-quality and single-crystal *wz*-ZnO SCs on Pt/Co$_{0.30}$Pt$_{0.70}$ electrodes and back-to-back Schottky diodes (BBSD) by MBE technique using Cu and Pt top electrodes. We report two main achievements: (1) fabrication of ZnO-SC with a FM contact with a high SBH of 600 meV, and (2) finding of a correlation between SBH and WF of metals and the extraction of a reasonable FLP energy.

## 2) Experimental methods

Samples of this report have the following structure: c-plane Al$_2$O$_3$ substrate/Pt (30, 700)/Co$_{0.30}$Pt$_{0.70}$ (10, 500)/ZnO (100, 400) [stack-A], c-plane Al$_2$O$_3$ substrate/Pt (30, 700)/Co$_{0.30}$Pt$_{0.70}$ (10, 500)/ZnO (130, 400)/Pt (20, 25) or Cu (200, 25) [stack-B], where the numbers in parentheses are layer thicknesses in nm and growth temperature in °C, respectively. The Al$_2$O$_3$ substrate was thermally annealed at 700 °C for 10 min to enhance the flatness of the surface. The metallic electrodes of Pt and Co$_{0.30}$Pt$_{0.70}$ were grown at a deposition rate of 1.2–1.8 nm/min, respectively, using a magnetron co-sputtering system. Then, ZnO thin films were deposited by MBE technique at 400 °C using an optimized O$_2$ flow of 2.0 SCCM (cubic centimeter per minute at STP) to maintain the magnetic properties of Co$_{0.30}$Pt$_{0.70}$ electrodes. The MBE chamber is accompanied with Zn and Mg Knudsen cells of high purity materials (4N) and attached to the co-sputtering chamber with an ultra-vacuum channel. The deposition rate of ZnO thin films was in the range of 4.2–5.4 nm/min. To enhance the quality of the top Pt/ZnO SC, ZnO thin films were post-treated by O$_2$ plasma for 120 sec at 200 °C prior to the Pt deposition, while the Cu deposition was done at RT to avoid its oxidation. The surface treatment of the Co$_{0.30}$Pt$_{0.70}$ was omitted to maintain the crystalline order and magnetic properties.

The in-situ and ex-situ surface morphologies, crystalline properties, magnetic properties of these stacks were investigated by reflection high-energy electron diffraction (RHEED) and atomic force microscopy (AFM), in-plane and out-of-plane X-ray diffractometer (XRD) (Bruker D8 Discover), and vibrating-sample magnetometer (VSM), respectively. The heterostructures crystalline quality was characterized using high-resolution transmission electron microscopy (HR-TEM). After deposition, samples of stack-B were patterned into a current-perpendicular-to-plan (CPP) circular devices of 100–300-µm diameters using dry etching, passivation of SiO$_2$ insulation and lift-off techniques (see **Fig.3 (b)**). The microfabrication was designed to minimize the effect of ZnO side milling and its degradation. The BBSD diodes were characterized by the conventional two-probes current–voltage (*I*–*V*) measurements where the voltage bias was applied to the top electrode using a picometer instrument of Keithley 6487.

### 3) Structural and magnetic properties

The RHEED images of c-sapphire substrate after thermal annealing, Pt (Co$_{0.30}$Pt$_{0.70}$) and ZnO surfaces after deposition are illustrated in **Figures 1 (a)**, **(b)**, **and (c)**, respectively. While **Figure 2 (a)** exhibited Kikuchi lines along azimuth [10–10], a clear indication of the high-quality single crystalline, Co$_{0.30}$Pt$_{0.70}$ and ZnO along azimuths [110] and [11–20] showed almost similar streaky lines indicating flat surfaces of these materials. Here, the surface quality of Pt buffer is similar to that of Co$_{0.30}$Pt$_{0.70}$ because of the proximity of their crystal parameters. The *ex-situ* surface morphologies of c-sapphire substrate, Pt and Co$_{0.30}$Pt$_{0.70}$ are illustrated in **Fig. 2 (d)**, **(e)**, **and (f)**, respectively. The Pt surface, with a roughness average (Ra) of 0.26 nm *vs.* 0.18 nm for the flat sapphire substate, exhibited a hexagonal spiral mode with steps reflecting the 2D growth mode and consistent with the small lattice mismatch between Al$_2$O$_3$(0001) and Pt(111) estimated to be 0.86%.[27] The Co$_{0.30}$Pt$_{0.70}$ surface is also smooth (Ra = 0.39 nm) as a bottom electrode but less than the Pt layer due to the CoPt alloy crystallization. The top-side AFM images of ZnO film is flat with an Ra of 0.34 nm, as shown in **Figure 1 (g)** consistent with the RHEED patterns and the small lattice mismatch between ZnO(0001) and CoPt(111) estimated to be 4.8%. High-resolution TEM images of a stack-A along Al$_2$O$_3$ [0001] axis is shown in **Figure 1 (h)**. All the stack-A composing layers were clearly distinguished and exhibited good crystallinity with abrupt interfaces. The absence of any element segregation or out-diffusion is confirmed within the enlarged yellow rectangle of **Figure 1 (i)** with some imperfections at the interface. In the latter one, the white rectangles corresponding to ZnO and $Co_{0.30}Pt_{0.70}$ films exhibited clear atomic alignments of both hcp-(0001) and fcc-(111), respectively. The blue and red circles inside the ZnO area exhibited the hcp classical zigzag pattern of Zn and O atoms as assigned by the violet line, while the black and gray circle inside CoPt area represented Co and Pt atoms of a well-textured $CoPt_3$-like alloy. Consequently, RHEED, AFM and HR-TEM images confirmed the deposition of high-quality Pt-based ferromagnetic electrode and ZnO thin film using a combination of co-sputtering and MBE deposition techniques and post-thermal treatments.

The crystalline properties of stack-A are shown in **Figure 2 (a)**. In the out-of-plane ω–2θ-scans, around Pt(111) and $Co_{0.30}Pt_{0.70}$(111) diffraction peak satellites due to Laue oscillations were observed, thus the achievement of coherent growth and high-quality thin films with a small interface roughness. *wz*-ZnO(0002) peaks were sharp without any additional phases. The other small peaks were confirmed to result from Cu-$K_β$ or W-$K_{α,β}$ radiations. In addition, the spectra of the in-plane φ-scans around asymmetric reflections of stack-A materials (**Figure 2 (b)**) exhibited a six-fold in-plane symmetry with sharp peaks for all layers with a clear epitaxial relationship: $Al_2O_3$(0001)[30–30]/ Pt(111)[220]/ $Co_{0.30}Pt_{0.70}$(111)[220]/ ZnO(0001)[11–20]. It is worth mentioning that we experimentally checked the ZnO film deposition on different 10–25-nm-thick $Co_{0.75}Pt_{0.25}$, $Co_{0.50}Pt_{0.50}$, and $Co_{0.30}Pt_{0.70}$ electrodes and found that the latter one, which a Pt-rich alloy, is more resistive to interface oxidation and allows *wz*-ZnO single-phase crystallization at moderate deposition temperatures.[27,45]

Concerning the magnetic properties of the bottom Pt/$Co_{0.30}Pt_{0.70}$ electrode, we did not detect neither a magnetization decay of a 10-nm-thick $Co_{0.30}Pt_{0.70}$ film between 10 K and 300 K, nor a magnetization degradation after the deposition of *wz*-ZnO film using MBE technique (**Figure 2 (c)**). The surface magnetic anisotropy energy of a 10nm-thick $Co_{0.30}Pt_{0.70}$ film was estimated to be +1.6 erg/$cm^2$ that is consistent with previous reports.[38,49] On the other hand, the magnetization dropped by ~20 % for a ZnO thin film deposited by the co-sputtering technique resulting in the formation of an interfacial CoO layer as shown in **Figure 2 (d)**. This was confirmed by the presence of elements intermixing at the FM/oxide interface and the broadening of the ZnO(0002) diffraction peak in the out-of-plan XRD spectrum (FWHM value of 0.16° for MBE vs. 0.45° for co-sputtering method, both aren't shown here). The MBE deposited $Co_{0.30}Pt_{0.70}$/*wz*-ZnO interface was smooth, well-ordered, and without any intermixing or multi-domains suggesting the superiority of MBE over co-spattering technique in fabricating high-quality *wz*-ZnO thin films on FM electrodes due to its lower deposition energy.[33,45] Hence, the MBE technique allowed high-quality and single-crystal *wz*-ZnO films deposition on a Pt/$Co_{0.30}Pt_{0.70}$ FM buffer at a moderate temperature of 400–450 °C.

## 4) Properties of ZnO-based Schottky Barriers

Since the semiconducting ZnO sandwiched between $Co_{0.30}Pt_{0.70}$ and Cu(Pt) is thick enough to avoid direct tunneling,[47] we can assume that the junction forms a BBSD with SBHs of $\Phi_{B1}\ and\ \Phi_{B2}$ as shown in the diagram of **Figure 3 (a)**. $\Phi_{B1}$ ($\Phi_{B2}$) is the SBH at $Co_{0.30}Pt_{0.70}$ and Cu(Pt) interfaces. In order to study the ZnO-based BBSDs, an equivalent circuit of **Figure 3 (b)** was adopted where two diodes D1 and D2 are connected in the back-to-back configuration. When $V_A$ is applied, one of the diodes is reversed-biased, while the other one is forward-biased. The total current will be limited by smaller current in the reverse biased diode. Here, $R_S$ is the series resistance resulting from the two ZnO SCs and ZnO thin film. $R_S$ value can be determined from the linear part of the *I-V* curve at high negative bias, i.e., $R_S \sim \Delta V_A/\Delta I$.[50] Actually, the voltage drop $IR_S$ needs be excluded from $V_A$ to obtain the effective voltage drop across D1 and D2. For a low turn-on voltage of a Schottky junction, it can be assumed that most of the voltage is distributed to the reversed-bias junctions. Finally, considering that the carrier transport across the junctions is dominated by the thermionic transport theory, the BBSD current of the equivalent circuit in **Figure 3 (b)** can be expressed by the following equations.[51]

$$I \approx \begin{cases} I_{S2}\ exp\left(-\frac{q(V_A-IR_S)}{n_2 k_B T}\right)\left[exp\left(\frac{q(V_A-IR_S)}{k_B T}\right)-1\right] & if\ V_A > 0 \\ I_{S1}\ exp\left(+\frac{q(V_A-IR_S)}{n_1 k_B T}\right)\left[1-exp\left(-\frac{q(V_A-IR_S)}{k_B T}\right)\right] & if\ V_A < 0 \end{cases} \quad (1)$$

$I_{S1(2)}$ is the reverse saturation current, $n_{1,2}$ is the ideality factor, *T* is the temperature in Kelvin, *q* is the elementary charge and $k_B$ is the Boltzmann constant. Numbers of 1 and 2 in the subscript of $I_S$, $\Phi_B$ and *n* refer to the corresponding junction D1 and D2, respectively. Therefore, Eq. (1) can be rewritten in the following forms.

$$\ln\left\{\frac{I}{exp\left(\frac{q(V_A-IR_S)}{k_BT}\right)-1}\right\} = \ln(I_{S2}) - \frac{q(V_A-IR_S)}{n_2k_BT} \qquad \text{for } V_A > 0 \qquad (2)$$

$$\ln\left\{\frac{I}{1-exp\left(-\frac{q(V_A-IR_S)}{k_BT}\right)}\right\} = \ln(I_{S1}) + \frac{q(V_A-IR_S)}{n_1k_BT} \qquad \text{for } V_A < 0 \qquad (3)$$

Plotting $\ln\{I/[exp(q(V_A - IR_S)/k_BT) - 1]\}$ versus $(V_A - IR_S)$ for positive $V_A$ results in a linear graph which the slope and y-axis intercept directly gives $n_2$ and $I_{S2}$, respectively. From the calculated $I_{S2}$, the SBH $\Phi_{B2}$ of D2 can be calculated:

$$q\,\Phi_{B1(2)} = k_BT\,\ln\left(S_{1(2)}\,A^*\,T^2/I_{S1(2)}\right) \qquad (4)$$

$S_{1(2)}$ is the SC area (circular of a diameter ϕ in the range of 100–300 μm) and A* is the ZnO effective Richardson constant (A* = 32 A cm$^{-2}$ K$^{-2}$ assuming m* = 0.27 m$_0$).[17] In the same way, the parameters of the junction D1 can be determined by utilizing Eq. (3) and (4).

The *I-V* curves of CPP pillars corresponding to applied voltages in the range of –2 to +2 V are shown in **Figure 3 (c)**. this range was selected to avoid breakdown of the devices measured at RT. Here, Cu, Co$_{0.30}$Pt$_{0.70}$ and Pt have a WF of 4.61 eV, 4.36 eV and 5.65 eV, respectively.[28,52] As a consequence, the absolute current for Pt top electrode is smaller than in the case of Cu electrode at negative bias voltages. The plotting of equations (2) and (3) for Cu topped Stack-B and the corresponding fitting parameters are illustrated in **Figure 3 (d)** and **(e)**. The extracted SBHs $\Phi_{B1}$ and $\Phi_{B2}$ of Co$_{0.30}$Pt$_{0.70}$/ZnO and Pt(Cu)/ZnO are summarized in **table .1**. Concerning $\Phi_{B1}$, the value for Cu/ZnO interface (613 meV) is smaller than that of the Pt/ZnO interface (690 meV) which is consistent with the higher WF of Pt compared to Cu. The value obtained from our fitting is consistent with the previous experimental reports of Pt/ZnO SC that ranged within 390–930 meV[17,32] and a limit of 950 meV from first principles calculations for Zn-terminated ZnO interface.[53] The Cu/ZnO SBH was close to the theoretical value of WF difference between Cu and ZnO (600 meV) and the recent experimental reports within 530–700 meV.[54,55] The value of $\Phi_{B2}$ at the Co$_{0.30}$Pt$_{0.70}$/ZnO interface for both top Pt and Cu electrodes took close values of 590 and 600 meV, that is lower than the Pt/ZnO SBH, reflecting the validity of the adopted fittings. While there is no report about the estimation of SBH at Co$_{0.30}$Pt$_{0.70}$/ZnO interface, the extracted $\Phi_{B2}$ value is bigger than the effective tunneling barrier height of 300 meV from our previous report [29,30] due to the suppression of the quantum tunneling in SCs. Therefore, the value of SBH at Co$_{0.30}$Pt$_{0.70}$/ZnO interface is a consequence of the optimized ZnO deposition by MBE which resulted in a high-quality interface as discussed in **section 3**. Here, the ideality factors $n_{1,2}$ of ZnO ranged within 1.06–1.22 which is close to the ideal value of 1, indicating that thermionic emission is the dominant transport phenomenon, as assumed in **Eq. (1)** to **(3)**. The slight shift can be associated to other phenomenon like interface states, generation-recombination assisted transport and the tunneling effect.[56] On the other hand, $R_S$ value for the BBSDs with the Cu and Pt top electrode is 105 kΩ and 285 kΩ, respectively. This is consistent with a SBH difference of ~90 meV and the $R_S$ proportional relationship to the exponential of $\Phi_B$, for a dominant Schottky transport.

Table. 1: The extracted parameters of ZnO-based BBSD using **Eq. (2)** and **(3)**.

| | | Top ZnO interface | | | $Co_{0.30}Pt_{0.70}$/ZnO interface | | |
|---|---|---|---|---|---|---|---|
| | | $\Phi_{B1}$ (meV) | ideality factor $n_1$ | $R_S$ (kΩ) | $\Phi_{B2}$ (meV) | ideality factor $n_2$ | $R_S$ (kΩ) |
| electrode | Cu (100 nm) | 613 | 1.25 | 105 | 600 | 1.06 | 105 |
| | Pt (50 nm) | 690 | 1.22 | 285 | 595 | 1.07 | 285 |

**Figure 4 (a)** summarizes the extracted SBHs metal/ZnO-SCs for Cu, $Co_{0.30}Pt_{0.70}$ and Pt contact metals in addition to the reported values with Ir, Pd, Au and Cu as a function of their WF values. Here, the degree of Fermi-level pinning (FLP) is characterized by the pinning factor $S$ $(= d\Phi_B/d\Phi_M)$, a parameter that depends on the density and extent of the interface states in the insulator, which is the slope of the dashed blue line. In general, $S$ varies from 0 for completely pinned states (Bardeen limit) to 1 for unpinned interfaces (Schottky–Mott limit, **Figure 4 (a)**). For the metal/ZnO SCs in this study, the value of SBH depends on the metal WF and the SBH increases for higher WF values. While our data points lie within the range of the experimental SBH of ZnO SCs from previous reports, the slop line is very close to the theoretical data of Ir/ZnO and in between those of Pd/ZnO and Au/ZnO.[17,28] The calculated value of $S$ is 0.08 which is much smaller than 1, indicating a strong FLP degree. While a systematic experimental study of the FLP at ZnO SCs isn't yet available, the $S$ value was theoretically estimated to range within 0.52–0.57 for free-defects metal/ZnO interfaces, for both polar and non-polar surfaces.[28,53,57] These values agree with the empirical formula of Mönch[57] using the dielectric constant of ZnO and were attributed to the presence of MIGS within the bandgap of ZnO, where the orbital of O at the SC dominantly contributes to their formation. Although MIGS are found at ZnO interfaces, it is thought that the pinning effect of MIGS is modest and that crystal defects are mostly responsible for the experimentally observed pinning[31,58]. In addition to the interfacial non-uniformity in epitaxial CoPt/ZnO SC, the existence of oxygen defects in our samples due to the medium temperature (400 °C) of ZnO deposition might explain the large reduction of $S$ factor from the theoretical predictions. Similarly, the degree of oxygen vacancies of epitaxial (001) and (010) β-$Ga_2O_3$ SC s plays an important role in the large reduction of the $S$ factor from the maximum value[59,60].

In order to describe the FLP effect, we adopted the (MIGS) model based on the charge neutrality level (CNL) and the pinning factor $S$.[28,61] **Figure 4 (b)** shows the band diagram of the metal/ZnO interface and metals where different energy levels are referenced to the vacuum level. According to the Schottky−Mott rule, SBH for the electron ($\Phi_B$) is given by the difference between the WF of a metal ($\Phi_M$) and the electron affinity of ZnO ($\chi_S$=4.30 eV)[62] when $S$ = 1; $\Phi_B = \Phi_M - \chi_S$. However, interface states aren't avoidable and get formed between the metal and ZnO which deviate the SBH from the Schottky−Mott rule, resulting in FLP effect. Here, the electron SBH is characterized quantitatively by implementing a pinning factor ($S$) and charge neutrality level (CNL, $\Phi_{CNL}$);

$$\Phi_B = S \cdot (\Phi_M - \Phi_{CNL}) + (\Phi_{CNL} - \chi_S) \qquad (5)$$

As well known, FLP effect is highly related to the defect-induced gap states of the semiconductors, the additional chemical disorders involved by device fabrications, and the MIGS of an energy level of $\Phi_{CNL}$.[61] At the metal/ZnO SC, the charge exchange between the MIGS on the ZnO side and metal; produces a dipole which tends to align CNL with the metal Fermi level and results in lowering $\Phi_B$. The extracted value of $\Phi_{CNL}$ using **Eq. (5)** is 4.94 eV, which lies 0.64 eV bellow the conductance band[26], *versus* 4.43 eV estimated from a theoretical study by Robertson and Falabretti[52] (3.27 eV above the valence band assuming a ZnO bandgap of 3.4 eV). The possible origins of MIGS at ZnO interfaces in this study are thought to be: (1) oxygen interstitials or singly (doubly) charged oxygen vacancies states[10] with an energy of 0.7 ± 0.2 eV below the conductance band,[31] consistent with the experimentally extracted $\Phi_{CNL}$ energy level, that is due to the ZnO natural vacancies at metals interfaces, (2) the modulation of DOS and spin-orbit coupling at the $Co_{0.30}Pt_{0.70}$ (Pt) interface by *3d-5d* hybridization driven by ZnO spontaneous polarization as demonstrated in our previous report.[29]

## 5) Conclusions

We succeeded in the deposition of high-quality ZnO-based BBSDs by MBE technique using an optimized $Pt/Co_{0.30}Pt_{0.70}$ bottom contact. While RXD measurements proved the single - phase crystallinity, HR-TEM showed the formation of abrupt interfaces. Also, the conservation of the $Pt/Co_{0.30}Pt_{0.70}$ magnetization; hence the absence of interface oxidation of Co; after ZnO deposition was confirmed by VSM. From *I–V* characteristics, we deduced reasonable SBHs of Pt/ZnO and Cu/ZnO interfaces, and a high SBH of 690 meV for the $Co_{0.30}Pt_{0.70}$/ZnO interface for the first time, up to our knowledge. The FLP effect was investigated by the contact metals WF dependence of the experimental SBHs. We obtained a pinning factor and a $\Phi_{CNL}$ level of 0.08 and 4.94 eV, respectively, indicating a strongly pinned Fermi-level by an MIGS that lies 0.64 eV bellow the conductance band. Hence, we confirmed that ZnO-SCs with metals grown by MBE technique follows the MIGS-model for SBHs as previously predicted, and the CNL might come from the oxygen interstitials or singly (doubly) charged oxygen vacancies states. These results imply the influence of defect states caused by atomic vacancies at SCs of oxide semiconductors. (total = 3430 words)

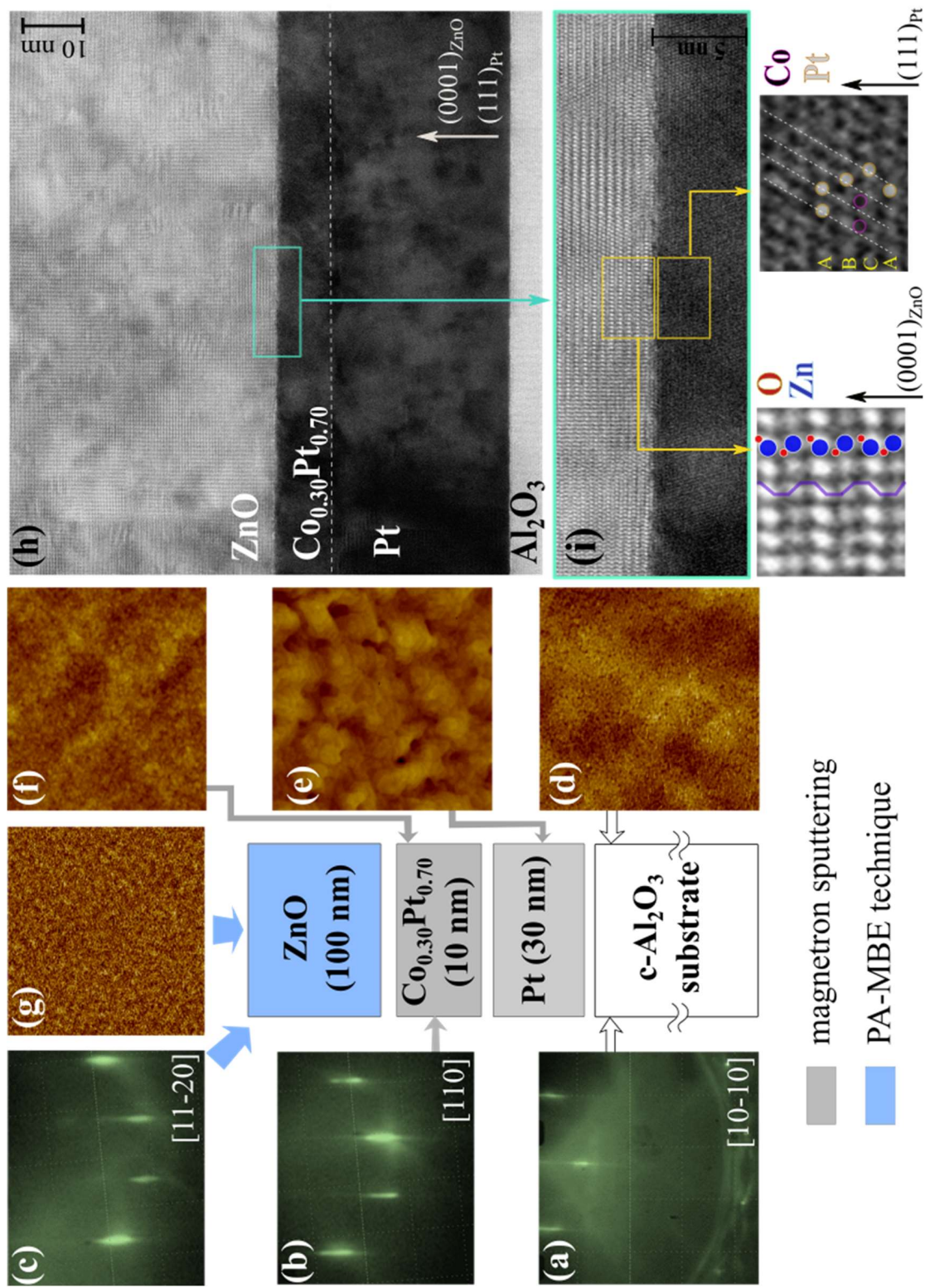

**Figure 1:** (a) and (d), (b) and (f), and (c) and (g) are the 1 × 1 μm² AFM images and *in-situ* RHEED patterns at the end of the deposition of c-Al$_2$O$_3$ substrate, Co$_{0.30}$Pt$_{0.70}$ (Pt) buffer, and ZnO epilayer, respectively. (e) 1 × 1μm² AFM scan of the 30-nm-thick Pt buffer surface. (h) and (i) are the high-resolution transmission electron microscopy (TEM) images of stack-A along [0001] axis and an enlarged region at the Co$_{0.30}$Pt$_{0.70}$/ZnO interface as indicated by the yellow rectangle. The enlarged regions of (i) shows the *wz*-ZnO(0001) zigzag and fcc-Co$_{0.30}$Pt$_{0.70}$ (111) atomic alignment.

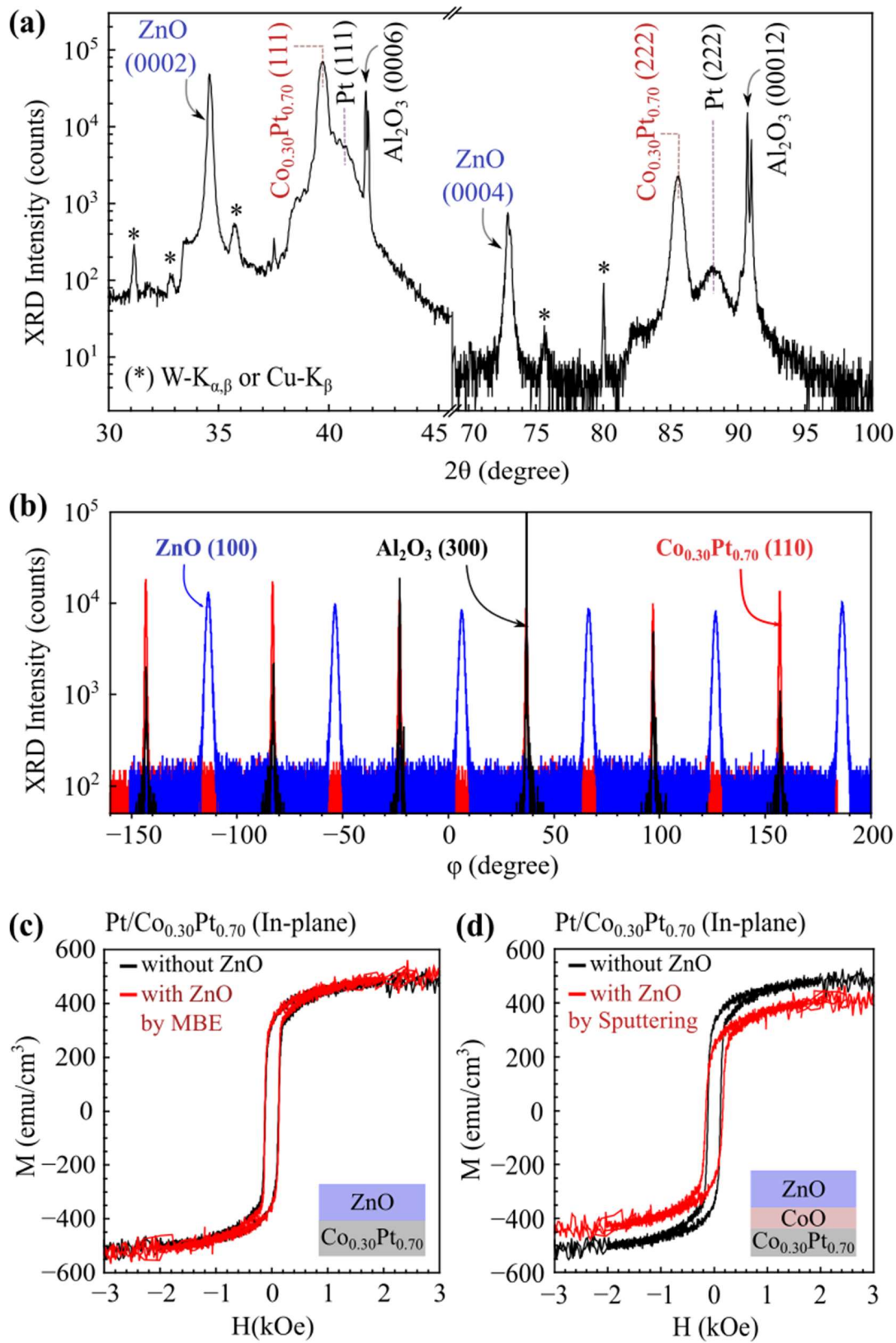

**Figure 2:** (a) ω-2θ XRD spectra of the c-Al$_2$O$_3$/Pt/Co$_{0.30}$Pt$_{0.70}$/ZnO structure (stack-A). (b) in-plane φ scan showing six sharp and periodic peaks of ZnO, Co$_{0.30}$Pt$_{0.70}$ and Pt materials. (c) and (d) are the in-plane VSM measurements of Pt/Co$_{0.30}$Pt$_{0.70}$ bilayer with and without the ZnO epilayer fabricated by MBE and sputter techniques, respectively.

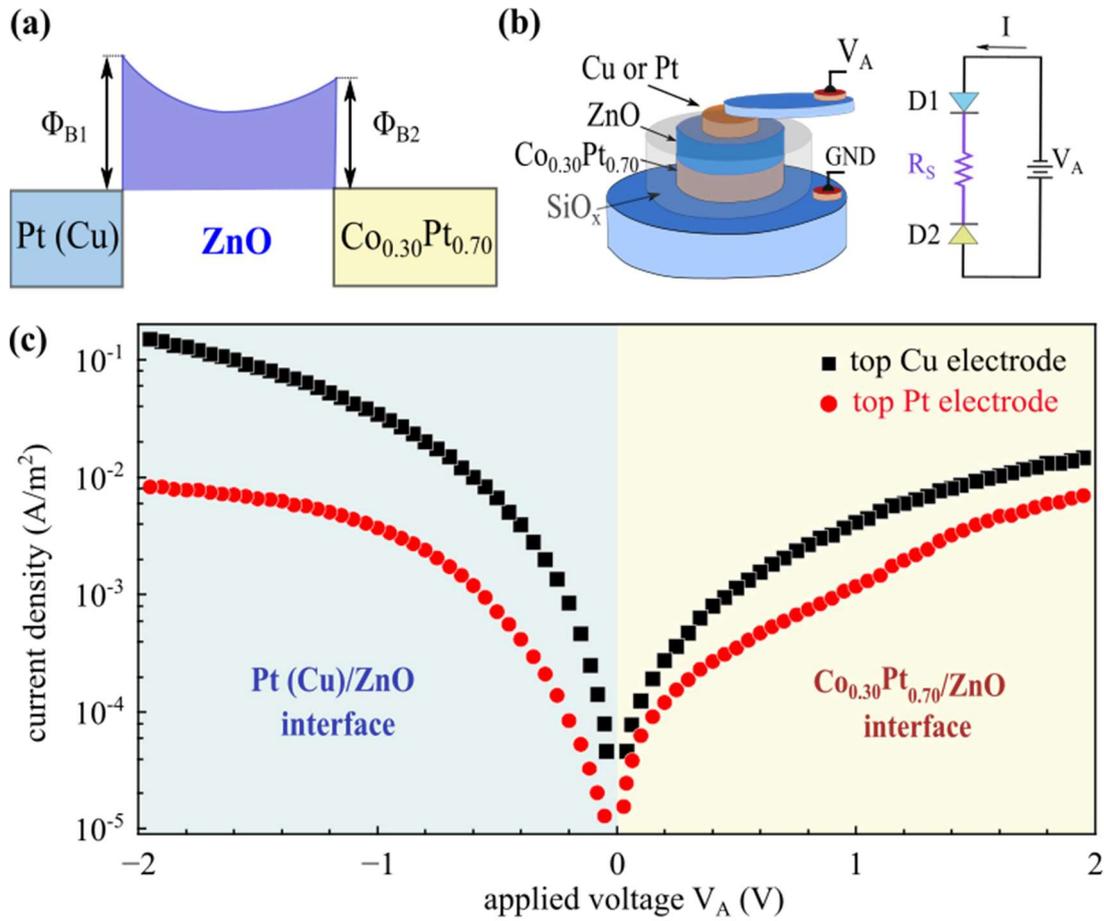

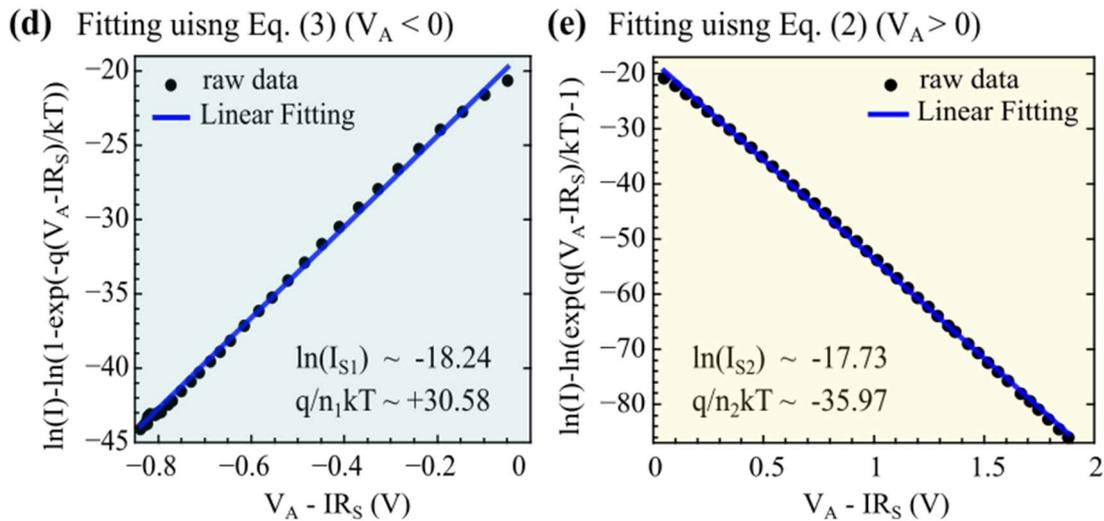

**Figure 3:** (a) The energy band diagram of ZnO-BBSD with the bottom $Co_{0.30}Pt_{0.70}$ and the top Pt(Cu) interfaces. $\Phi_{B1}(\Phi_{B2})$ is the corresponding SBH. (b) A representation of the patterned CPP device used for the *I-V* curves measurements (left), and a back-to-back configuration of two ZnO-Schottky barriers connected with series resistance $R_S$ (right). (c) *I-V* curves of two ZnO-based BBSDs with Cu and Pt top electrodes. The corresponding ZnO interfaces are indicated within the graph. (d) and (e) are the *I-V* data fittings for the extraction of $\Phi_B$ and n using equations (3) and (2) for Pt(Cu)/ZnO and $Co_{0.30}Pt_{0.70}$/ZnO interfaces. The extracted fitting parameters are added inside the figures while the ZnO-SCs parameters are summarized in table. 1.

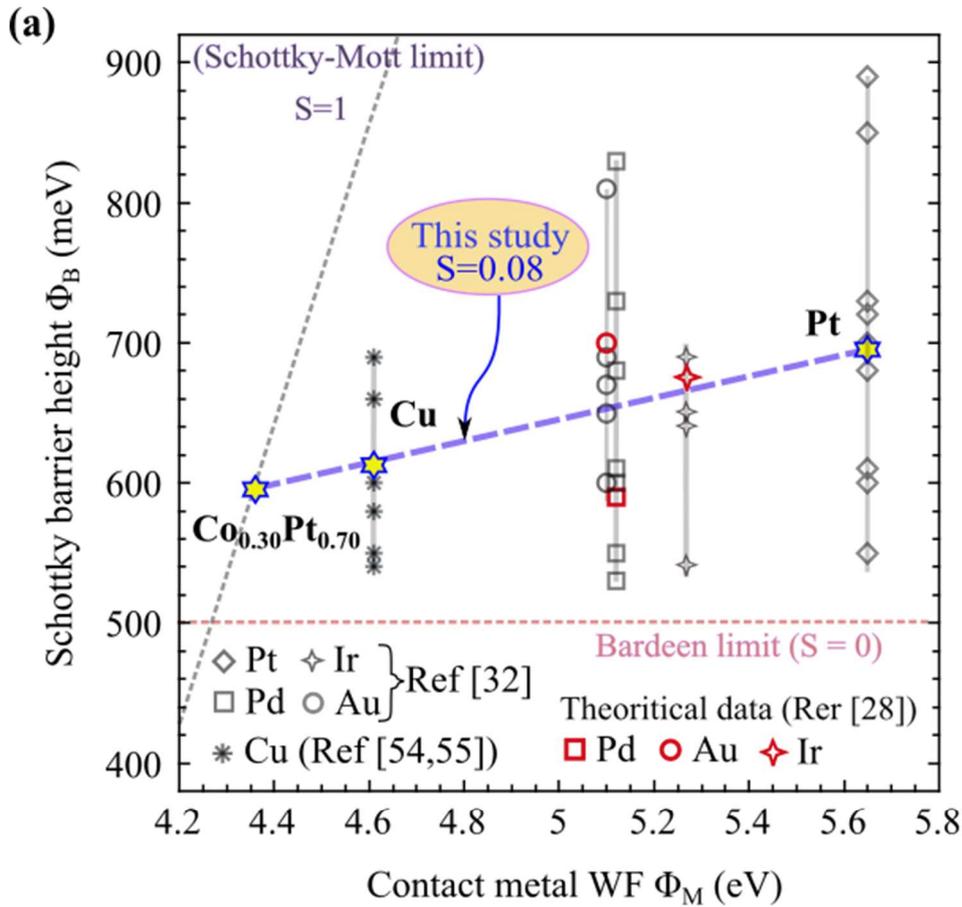

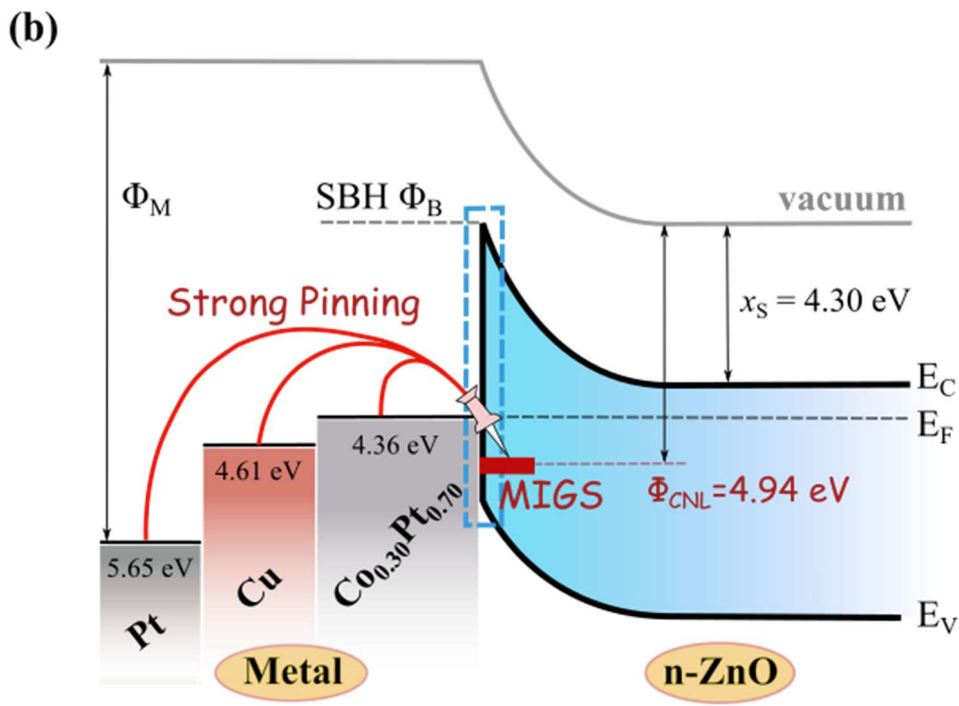

**Figure 4:** (a) SBH of ZnO for the various metal work functions. The pinning factor S were obtained from fits using the Schottky–Mott rule (dotted violet line). Both Schottky–Mott (S=1) and Bardeen limits (S=0) were added for comparison. SBHs of ZnO with Cu, Pt, Ir, Pd, Au metals were also added.[32,55] (b) Band diagram at the contact with ZnO including defect states (MIGS), metals, and $\Phi_{CNL}$ states. These can change the SBH and create a strong Fermi-level pinning by $\Phi_{CNL}$.

**Author Information**

Mohamed Belmoubarik - International Iberian Nanotechnology Laboratory, INL, Av. Mestre José Veiga s/n, 4715-330, Braga, Portugal.

orcid.org/0000-0003-3592-1259;    Email: mohamed.belmoubarik@inl.int

**Data Availability.**

The data that support the findings of this study are available from the corresponding Authors upon reasonable request.

**Declaration of Competing Interest.**

The authors declare that they have no known competing financial interests or personal relationships that could have appeared to influence the work reported in this paper.

**Acknowledgment.**

This work was sponsored by Japan Society for the Promotion of Science (Grant No. 25-5806). We are grateful to Professor Masashi Sahashi, Dr. Tomohiro Nozaki, Mr. Hideyuki Sato of Tohoku University and Dr. George Junior of INL Portugal for their valuable technical support and fruitful scientific discussions. The author is also grateful to his wife Fadwa Mhamdi Alaoui for the full support during the draft preparation.